# Asymmetric depinning of chiral domain walls in ferromagnetic trilayers

Mark C. H. de Jong[1,2], Can Onur Avci[1], Aleš Hrabec[1,3], and Pietro Gambardella[1]

[1]*Department of Materials, ETH Zürich, CH-8093 Zürich, Switzerland*

[2]*Department of Applied Physics, Eindhoven University of Technology, P.O. Box 513, 5600 MB Eindhoven, the Netherlands*

[3]*Paul Scherrer Institut, Villigen, Switzerland*

**We show that the coupling between two ferromagnetic layers separated by a nonmagnetic spacer can be used to control the depinning of domain walls and induce unidirectional domain wall propagation. We investigated CoFeB/Ti/CoFeB trilayers where the easy axis of the magnetization of the top CoFeB layer is out-of-plane and that of the bottom layer is in-plane. Using Magneto-optic Kerr effect microscopy, we find that the depinning of a domain wall in the perpendicularly magnetized CoFeB layer is influenced by the orientation of the magnetization of the in-plane layer, which gives rise to a field-driven asymmetric domain expansion. This effect occurs due to the magnetic coupling between the internal magnetization of the domain wall and the magnetization of the in-plane CoFeB layer, which breaks the symmetry of up-down and down-up homochiral Néel domain walls in the perpendicular CoFeB layer. Micromagnetic simulations support these findings by showing that the interlayer coupling either opposes or favors the Dzyaloshinskii-Moriya interaction in the domain wall, thereby generating an imbalance in the depinning fields. This effect also allows for artificially controlling the chirality and dynamics of domain walls in magnetic layers lacking a strong Dzyaloshinskii-Moriya interaction.**

**I - INTRODUCTION**

The development of future magnetic memory and logic technologies requires efficient control of magnetization in nanometer-sized devices[1]. In a ferromagnet (FM), information can be encoded in magnetic domains and manipulated by displacing them[2,3]. A variety of device concepts based on domain wall (DW) motion has been proposed for memory and logic operations[4-6]. Many of these proposals rely on unidirectional DW motion along a racetrack, which can be easily achieved by current-induced spin-torques[7-16], rather than externally applied magnetic fields[3].

In layered heterostructures, the interaction between neighboring magnetic layers strongly influences the behavior of the individual FMs. A prominent example of interlayer interactions is the Ruderman-Kittel-Kasuya-Yosida (RKKY) coupling between two FMs separated by a nonmagnetic spacer, which stabilizes either the parallel or antiparallel alignment of the magnetization vectors in neighboring FMs[17,18]. The RKKY interaction has been key to the discovery of the giant magnetoresistance[19,20] and has remained an active research topic ever since[21-25]. In combination with the Dzyaloshinskii-Moriya interaction (DMI), the RKKY interaction is responsible for chiral interlayer coupling[26,27], which has recently attracted renewed attention[28-30]. Other interlayer coupling mechanisms, such as the direct exchange bias between FMs and antiferromagnets[31,32] and interlayer dipolar interactions[33-35], are relevant for tuning the response of magnetic tunnel junctions and spin valves to external fields and currents. The rich physics and the applications enabled by interlayer coupling effects in magnetic heterostructures are highly attractive for research purposes. However, with few exceptions[21,22,36-42], the majority of such effects have been so far explored in uniformly magnetized layers.

In this article, we investigate the depinning of DWs in a FM layer with out-of-plane (OOP) magnetization coupled through a nonmagnetic spacer to a FM layer with uniform in-plane (IP) magnetization. When both layers are saturated along their respective easy axis there is no apparent effect on their response to external fields. Conversely, when a DW is nucleated in the OOP layer, its internal magnetization couples to the IP layer, lifting the energy degeneracy of up-down and down-up DWs. This coupling offsets the depinning field of up-down and down-up DWs and causes asymmetric field-induced expansion of domains in the OOP layer. Consequently, in an up-down-up domain configuration (three domains separated by two DWs), we are able to move only one of the DWs for a range of values of the OOP field. The moving DW is controlled by the magnetic polarity of the IP layer, which is set prior to the measurement. This effect opens up the possibility to realize unidirectional field-driven DW displacement without requiring a current-induced spin-torque mechanism[8,43] or an externally applied IP field during the displacement[44,45]. Moreover, this coupling effect can be used to artificially control the chirality and dynamics of DWs in magnetic layers lacking DMI[21].

## II - SAMPLE PREPARATION AND ELECTRICAL CHARACTERIZATION

We deposited //Ti(2)/CoFeB(3)/Ti($t$)/CoFeB(1)/MgO(2)/Ti(2) by magnetron sputtering onto a Si/SiO$_2$(300) substrate at room temperature. The numbers correspond to the thickness in nm. The Ti spacer thickness $t$ was varied between 1.5 and 4 nm. The composition of CoFeB was 40% Co, 40% Fe and 20% B. All metallic layers were grown by d.c. sputtering in a base pressure of $\sim 0.5 \times 10^{-8}$ mbar and Ar partial pressure of $2 \times 10^{-3}$ mbar. The MgO layer was deposited using r.f. sputtering. 5-μm-wide DW tracks have been defined on blanket substrates using UV photolithography. After deposition, the devices were obtained by lift-off. To promote the perpendicular magnetic anisotropy of the top CoFeB layer, after lift-off, the samples were annealed at 250 °C for 45 minutes in a vacuum chamber at a pressure <$10^{-6}$ mbar. Finally, 50-nm-thick and 3-μm-wide Au wires were deposited orthogonal to the DW tracks to controllably nucleate domains via the Oersted field produced by current pulses. A typical device is depicted in Fig. 1 (a). The thicker bottom CoFeB layer has IP anisotropy and the Ti spacer breaks the direct coupling of the two CoFeB layers[46], such that the top and bottom CoFeB magnetization direction can be independently controlled. For electrical characterization, we measured Hall effect and magnetoresistance using standard a.c. current injection methods. All experiments were performed in ambient conditions.

We first characterize the Hall effect and magnetoresistance in a typical device with a Ti(2 nm) spacer layer. For an OOP magnetic field ($B_z$) sweep, the Hall resistance ($R_H$) is proportional to the OOP magnetization component through the anomalous Hall effect (AHE), therefore we expect that the signal predominantly originates from the OOP CoFeB layer. Figure 1 (b) shows $R_H$ versus $B_z$, exhibiting a clear hysteresis with a coercivity of $B_c = 27 \pm 2$ Oe and 100% remanence at zero field, which is an indication of robust perpendicular magnetic anisotropy in the OOP CoFeB layer. Figure 1 (c) shows the longitudinal resistance ($R_L$) measured between one of the transverse contacts and the Au nucleation line [see Fig. 1 (a)] for applied fields along ($B_x$) and across ($B_y$) the DW track. Due to the anisotropic magnetoresistance, we expect $R_L$ to be maximum when the magnetization is parallel to the track ($x$-axis) and minimum when it is orthogonal ($y$-axis) to it[47]. $R_L$ is nearly constant over the entire field range when sweeping $B_x$, whereas it decreases significantly as the field is increased when sweeping $B_y$, saturating around $B_y \sim 60$ Oe. The $R_L$ measurements clearly indicate that, at equilibrium, the magnetization of the IP CoFeB layer points along the *x*-axis due to the relatively strong shape anisotropy (in excess of 50 Oe) induced by the DW track geometry.

## III - RESULTS AND DISCUSSION

### A. Asymmetric DW depinning experiments

To characterize the DW depinning properties, the same sample is mounted in a wide-field magneto-optic Kerr (MOKE) microscope in polar geometry which is sensitive to the OOP component of the magnetization and used to image the domains in the OOP CoFeB layer. By applying a sequence of external fields, $B_z = 150$ Oe followed by $B_x = 30$ Oe, the magnetization of the device is preset in the [→↑] configuration. Here the arrows indicate the magnetization direction of the IP and OOP CoFeB layers, respectively. We then nucleate a reversed domain in the OOP CoFeB by sending a current pulse through the Au wire and apply $B_x = 30$ Oe again for a short time, to reset any undesirable influence of the nucleation current on the IP CoFeB layer. The magnetization state in the system after the above procedure is schematically shown in Fig. 2 (a). When $B_z$ is swept (in the absence of $B_x$) towards negative values we observe that the down-up DW depins first, at $B_{\text{dep},\downarrow\uparrow} = -7.3 \pm 0.3$ Oe, then the up-down DW depins at a larger field $B_{\text{dep},\uparrow\downarrow} = -11.8 \pm 0.1$ Oe, as shown in Fig.2 (a) - bottom panel. Instead, if the magnetization of the IP CoFeB layer is saturated along the negative $-x$-direction, corresponding to the [←↑] initial state [Fig. 2 (b) - top panel], we observe that the asymmetry in the domain expansion is reversed, as the up-down DW depins at $B_{\text{dep},\uparrow\downarrow} = -7.4 \pm 0.4$ Oe and the down-up DW depins at $B_{\text{dep},\downarrow\uparrow} = -12 \pm 1$ Oe. Reciprocal processes are observed for the [→↓] and [←↓] initial states, as shown in Fig. 2 (c) and (d). Figure 2 (e) and 2 (f) report the field evolution of the asymmetric domain expansion for the initial configurations shown in Fig. 2 (a) and 2 (b), respectively. These measurements were repeated on ~10 different devices, prepared in two separate deposition runs, and we obtained the same qualitative behavior.

The dependence of the observed asymmetry on the orientation of the magnetization in the IP CoFeB layer reveals a magnetic interaction between the two layers that depends on the type of DW, namely up-down vs. down-up. In this respect, our observations differ from the previously reported asymmetry of the DW creep velocity in antiferromagnetically coupled Co/Pt/Co trilayers, where both Co layers have OOP magnetization and the interlayer coupling always favors the expansion of antiparallel domains[36,37]. There are two possible scenarios that could break the symmetry of the DWs and give rise to the effect reported here: *(i)* a magnetostatic coupling between the magnetization of the IP CoFeB and the internal magnetization of the DWs in the OOP CoFeB, e.g., via the Neel 'Orange Peel' effect[33-35,48] [see Fig. 3 (a)], and *(ii)* the RKKY coupling between the internal magnetization of the DWs in the OOP CoFeB and the magnetization of the IP CoFeB [see Fig. 3 (b)]. In both cases, our observations require the internal magnetization of the domain walls to be oriented in-plane, parallel or antiparallel to the magnetization of the bottom CoFeB layer.

### B. Spacer layer thickness dependence of the asymmetric DW depinning

To test the hypotheses given in Sect. III-A, we measure the DW depinning asymmetry as a function of the thickness of the Ti spacer. The depinning asymmetry is quantified by $\Delta B_{\text{dep}} = B_{\text{dep,left}} - B_{\text{dep,right}}$, i.e., the difference in the depinning fields between the left and right DWs in an up-down-up or down-up-down configuration. Although the depinning fields can be device or sample-dependent, the differential measurement $\Delta B_{\text{dep}}$ should scale with the coupling strength and depend less on the device under consideration. Figure 3 (c) shows $\Delta B_{\text{dep}}$ as a function of the Ti thickness. We observe that the sign of the asymmetry is the same in each of the devices with different Ti thickness, but its magnitude decreases as the Ti thickness is increased. The strong thickness dependence of $\Delta B_{\text{dep}}$ confirms that the magnetostatic coupling of the OOP domains to the magnetization of the IP layer is not responsible for the asymmetric DW expansion, since in this case the length scale on which the coupling should persist is expected to be on the order of the domain size itself (i.e., ~μm)[6,7]. The nearly four-fold decrease when increasing the Ti thickness from 1.5 nm to 4 nm suggests that the coupling occurs locally between the DWs and the bottom CoFeB layer.

A local coupling effect is expected to act as an effective field on the internal magnetization of the DWs, which can be either parallel or antiparallel to the magnetization direction of the bottom CoFeB layer depending on the origin of coupling. This effective field can significantly change the DWs depinning potential and velocity if the DWs are of homochiral Néel-type[45], similar to the bubble expansion experiments reported in Refs.[44,49,50]. If the DWs were entirely of Bloch type instead, the coupling with the IP layer pointing either → or ← could not generate any asymmetry between the energies of the right- and left-handed DWs.

## C. Determination of the domain wall chirality and Dzyaloshinskii-Moriya interaction in Ti/CoFeB/MgO layers

In thin films with OOP magnetization, the formation of chiral Néel DWs is induced by the Dzyaloshinskii-Moriya interaction (DMI). To show that the DWs are indeed chiral in our system, we characterized the DMI in a similar structure [substrate//Ti(2)/Pt(5)/Ti(2)/CoFeB(1)/MgO(2)/Ti(2)] *via* current-induced DW depinning experiments[15,51]. Note that here we used Pt instead of the bottom CoFeB in order to generate the necessary spin-orbit torque (SOT) to move the DWs[11,12] [Fig. 4 (a)], but due to the relatively thick Ti(2 nm) layer between Pt and CoFeB we do not expect Pt to induce a sizeable DMI on the OOP CoFeB[12,52]. To determine the equilibrium orientation of the DW moment in the CoFeB layer, we followed the experimental procedure previously reported by Avci et al.[15]. After nucleating a DW near the Au wire, we applied current pulses through the DW track, while simultaneously applying a magnetic field along the *x*-axis ($B_x$). By applying $B_x$ of different amplitudes, the orientation of the DW internal magnetization can be rotated to any angle in the *xy*-plane. Then the DMI effective field can be, in principle, estimated by the $B_x$ values at which the polarity of the depinning current (to move the DW along the same direction) changes sign, which indicates a chirality change in the DW. We recorded the current density required to depin the DW from its initial position for both DW types for -95 Oe < $B_x$ < 95 Oe.

Figure 4 (b) shows the results of the above experiments. At $B_x$ = 0, both up-down and down-up DWs require a negative current density, $j_d = (-1.0 \pm 0.1) \times 10^7$ A cm$^{-2}$, to move along the $-x$ direction. This is consistent with the SOT-driven DW motion along the current direction for chiral Néel DWs, whereas for the conventional spin transfer torque one expects the DWs to move in the opposite direction (electron direction). Upon application of a negative (positive) $B_x$ the current density required to depin the DW is further reduced for the up-down (down-up) DWs. This means that at zero field the DW is in between Néel and Bloch-type, and that a negative (positive) $B_x$ rotates the internal magnetization of up-down (down-up) DWs towards the Néel configuration favored by the DMI effective field. On the other hand, application of a positive (negative) $B_x$ increases the current density required to depin the DW for the up-down (down-up) DWs up to $|B_x| \sim 30$ Oe, after which the DW becomes fully of Bloch-type. At larger fields, the depinning current first stabilizes and then changes sign (gray shaded areas). The sudden sign reversal of $j_d$ reflects the chirality change of the DWs induced by the larger $B_x$ applied opposite to the DMI effective field. These data unequivocally show that the DWs in the Ti/CoFeB/MgO tracks have left-handed chirality with a DMI effective field of the order of 30±10 Oe. The DMI constant can then be calculated by $D = \mu_0 H_{\text{DMI}} M_s \Delta$ (Ref. 53), where $M_s = 1 \times 10^6$ A m$^{-1}$ is the saturation magnetization and $\Delta_{\text{DW}} = 20 \pm 5$ nm is the domain wall width determined for similar Ta/CoFeB/MgO structures[54]. This gives $|D| = 0.06 \pm 0.02$ mJ m$^{-2}$. This result is consistent with recent reports that a DMI can be induced at FM/oxide interfaces without requiring a heavy metal layer[55,56].

### D. Discussion of the different coupling scenarios

The results presented above show that the DWs in the OOP CoFeB layer have a chiral Néel character, hence IP internal DW magnetization couple to the IP magnetization of the bottom CoFeB layer. This coupling generates an imbalance in the energy of up-down and down-up DWs, which have opposite IP magnetization components. As a result, the pinning potential of the DWs depends on the relative alignment of the IP effective field and internal DW magnetization. Henceforth, we discuss the origin of the interlayer coupling. In scenario *(i)*, given the strong dependence of the depinning field on the Ti spacer thickness, a magnetostatic interaction between the DWs in the top CoFeB layer and the magnetization of the bottom CoFeB layer can only happen due to the 'Orange Peel' effect[33-35,48] as depicted in Fig. 3 (a). If the interface between the bottom CoFeB layer and the Ti spacer is rough enough, the magnetic surface charges will generate a magnetic field parallel to the magnetization of the IP CoFeB layer, which can couple to the magnetization in the OOP CoFeB layer. A recent study showed that in CoFeB(3)/W/CoFeB(1)/MgO(2) structures this effect can produce a sizeable magnetic field (~10 Oe) acting on the top CoFeB layer for a W thickness up to 4 nm[57].

In scenario *(ii)* an electronic coupling between the DWs and the IP CoFeB can be mediated by the RKKY interaction [Fig. 3 (b)]. In the uniformly magnetized state of both CoFeB layers, the RKKY will have a negligible influence on the magnetic behavior of the individual layers since the exchange energy is proportional to the dot product of the magnetization vectors. However, the RKKY is nonzero when the OOP CoFeB has an IP magnetization component such as in a DW. The effect of the coupling is analogous to that of an effective field along the *x*-direction, given by $H_{\text{RKKY}} = 2J/(M_s t_{\text{FM}})$, where $t_{\text{FM}}$ is the thickness of the FM layer, $M_s$ the saturation magnetization, and $J$ is the coupling strength, whose sign and magnitude oscillate with the thickness of the spacer layer with a period of ~1 nm[18]. The absence of noticeable oscillations of $\Delta B_{\text{dep}}$ as a function of Ti thickness [Fig. 3(c)] indicates that the RKKY interaction is not responsible for the asymmetric DW depinning field. Indeed, previous studies showed that Ti is a poor mediator of the RKKY coupling compared to other 3d transition metals such as Cr or V (see Ref. [17]). By taking $\Delta B_{\text{dep}}$ as a measure of the coupling magnitude, if the coupling were RKKY-mediated, it should be on the order of a few Oersted, corresponding to $J \sim 0.4 \times 10^{-3}$ erg/cm$^2$ (assuming an exchange field of $H_{\text{RKKY}}$ = 3 Oe, $M_s$=1100 emu/cm$^3$ and $t_{\text{Ti}}$ = 2 nm). An RKKY effect of such a magnitude is about 3 orders of magnitude smaller than reported for the standard Co/Ru/Co multilayers[58]. Therefore, the origin of the asymmetry in the DW depinning and expansion is ascribed to a magnetostatic orange-peel effect [scenario *(i)*], although minor contributions from the RKKY coupling cannot be excluded.

### E. Micromagnetic simulations

We further performed micromagnetic simulations using MuMax3 (Ref. [59]), in order to evaluate the effect of a magnetic coupling between the DWs and the IP CoFeB layer on the depinning field $B_{\text{dep}}$. As noted previously, the effect of the magnetic coupling on the DW is an effective field along the magnetization direction of CoFeB (either along +*x* or -*x*), therefore we have used a static IP field ($\pm B_x$) in order to simplify the simulations instead of the IP CoFeB layer. Here the amplitude of $\pm B_x$ qualitatively reflects the thickness variation of Ti, i.e., larger (smaller) $\pm B_x$ in the simulations corresponds to thinner (thicker) Ti in the experiments. The magnetic defect from which the DW depins was modelled by a 50 nm-wide region where the strength of the anisotropy and DMI are slightly increased. The simulated magnetic layer has a saturation magnetization $M_s$ = 1x10$^6$ A/m, interfacial DMI energy $D$ = 0.1 mJ/m$^2$, exchange stiffness constant $A$ = 16 pJ/m, and perpendicular magnetic anisotropy constant $K_u$=8.78x10$^5$ A/m, as expected for the OOP CoFeB layer studied here[60]. At the defect, $K_u$ and $D$ were increased to 8.98x10$^5$ A/m and 0.11 mJ/m$^2$, respectively, since these values are known to scale with the thickness of the FM layer[61,62]. As derived analytically in the Appendix, the modification of $D$ at the pinning potential was crucial in obtaining the DW depinning asymmetry for chiral DWs.

Figure 5 (a) shows $B_{\text{dep}}$ (applied along the $\pm z$-axis) required to depin the DWs from the defect in the simulations as a function of a simultaneously applied $B_x$ for four different cases: up-down (solid symbols) and down-up (open symbols) DWs in the absence (red dots) and presence (blue squares) of the DMI. The absence of DMI results in achiral Bloch-type DWs, whereas the presence of (weak) DMI results in partially left-handed Néel-type chiral DWs, as experimentally determined in our samples. In the absence of DMI ($D = 0$), $B_{\text{dep}}$ is about 10 mT irrespective of the DW type and has a weak dependence on $B_x$. In the presence of DMI we observe that $B_{\text{dep}}$ becomes strongly dependent on $B_x$ (Ref. 45,63), and that $B_{\text{dep}}$ decreases for both DW types when $B_x = 0$. The decrease in $B_{\text{dep}}$ at $B_x = 0$ is the result of a decrease of the pinning potential due to the increase in DMI at the pinning site, as we illustrate in the Appendix.

Unlike the observed decrease of $B_{\text{dep}}$ at $B_x = 0$ mT, the dependence on $B_x$ is different for each DW type. For an up-down DW (solid blue squares), where the spin orientation in the center of the DW has a component along the $-x$ direction at equilibrium, $B_{\text{dep}}$ is low for $B_x < 0$ and gradually increases and saturates upon increasing $B_x$. This is because $B_x > 0$ acts against the DMI effective field, first aligning the spin orientation in the center of the DW with the y-axis (Bloch DW) around $B_x = 15$ mT, where $B_{\text{dep}}$ becomes equal to the $D = 0$ case, and then increasing further when the DW spin orientation rotates to the right-handed chirality. The saturation occurs at $B_x > 30$ mT as the spin orientation in the center of the DW is fully aligned with the external field and no further changes occur beyond this field. The energy of a DW in this regime depends on the sum of the DMI effective field and the applied in-plane field[44]. At the pinning site, the strength of the DMI effective field increases, which results in a different pinning potential if the two fields are parallel or antiparallel, as we show in the Appendix. The data for the down-up DW is mirror symmetric with respect $B_x = 0$ due to the reciprocity of the above described processes. Figure 5 (b) shows the difference in the depinning fields ($\Delta B_{\text{dep}}$) between the up-down and down-up DWs. As expected from Fig. 5 (a), $\Delta B_{\text{dep}}$ is nearly zero for the $D = 0$ case, whereas it has a strong $B_x$ dependence in the presence of DMI. These simulations show that the interplay of the DMI and the IP magnetic fields supports the asymmetry of the depinning fields observed experimentally.

## IV - CONCLUSIONS

In conclusion, we have investigated the field-induced DW depinning in CoFeB/Ti/CoFeB trilayers where the top CoFeB layer has OOP magnetization and the bottom CoFeB layer has IP magnetization. The DW depinning field varies by up to 50% for up-down and down-up DWs for a given polarization of the IP CoFeB. The DW depinning field asymmetry decreases monotonically as a function of the Ti thickness and almost vanishes for Ti(4 nm), demonstrating that it originates from the magnetostatic orange-peel coupling between the IP CoFeB and the internal magnetization of the DW in OOP CoFeB. The DWs in the OOP CoFeB have a left-handed Néel-type spin structure, with the internal magnetization pointing either → or ← for down-up and up-down DWs, respectively. The coupling with the magnetization of the IP CoFeB generates an imbalance in their internal energy, which offsets the field-induced depinning and expansion of down-up and up-down DWs. We validated these findings with micromagnetic simulations, which showed that the asymmetric domain expansion is due to the interplay between the DMI and the effective field generated by the IP layer on the DW spin orientation. Such an effect allows for achieving unidirectional domain displacement by carefully programming the sequence of IP and OOP fields. A simple model of the DW energy suggests that the asymmetry of the pinning potential is proportional to the local variation of the DMI (see Appendix). In general, our study shows that interlayer coupling can be used to artificially induce and control the type and chirality of DWs in magnetic layers lacking DMI.

**Acknowledgements**

This work was supported by the Swiss National Science Foundation through grant #200020_172775. A.H. was funded by the European Union's Horizon 2020 research and innovation programme under Marie Sklodowska-Curie grant agreement no. 794207 (ASIQS).

**APPENDIX: TOY MODEL OF THE DOMAIN WALL PINNING POTENTIAL IN THE PRESENCE OF DMI**

To illustrate how an applied in-plane field can affect DW pinning in the presence of DMI we present a toy model of the pinning potential at the simulated pinning site (see Sect. III-E). Note that this model is not comprehensive. Derivation of a complete analytical model of the depinning process is beyond the scope of this work. To a first approximation of the one-dimensional model investigated by micromagnetic simulations, the surface energy density of a DW can be written as[44],

$$\sigma_{DW} = \sigma_0 - \frac{\pi^2 \Delta_{DW} \mu_0^2 M_s^2}{8K_D}(H_x + H_{\text{DMI}})^2, \quad \text{(A1)}$$

where $\sigma_0 = 4\sqrt{AK_{\text{eff}}}$ is the Bloch DW energy density, $\Delta_{DW}$ is the DW width, $K_D$ is the DW anisotropy energy density, $H_x(= B_x/\mu_0)$ is the applied in-plane magnetic field, and $H_{\text{DMI}} \propto D$ is the DMI effective field. This approximation holds if the sum $H_x + H_{\text{DMI}}$ is not strong enough to force the DW to be of the Néel type. We model the pinning potential as a small increase in $K_{\text{eff}}$ and $H_{\text{DMI}}$, resulting in a different DW energy for the pinning site and the rest of the track. Using Equation (A1), we calculate this difference in the DW energy density, assuming that a small change in the DW width $\Delta_{DW}$ can be neglected and find,

$$\Delta\sigma_{\text{DW}} \approx \Delta\sigma_0 - \frac{\pi^2 \, \Delta_{DW} \, \mu_0^2 M_s^2}{8 K_D} \left[ 2 H_{\text{DMI}} \Delta H_{\text{DMI}} + 2 H_x \Delta H_{\text{DMI}} + \Delta H_{\text{DMI}}^2 \right], \quad (A2)$$

where $\Delta\sigma_0$ is the difference in the Bloch DW energy density and $\Delta H_{\text{DMI}}$ the difference in the DMI effective field. The first and third term in the square brackets change the pinning potential at $H_x = 0$. We confirmed that $B_{\text{dep}}$ scales approximately linearly with $\Delta H_{\text{DMI}}$ in the simulations (since $2 H_{\text{DMI}} > \Delta H_{\text{DMI}}$). For an up-down and down-up DW wall, the second term in the square brackets will have opposite sign. Increasing the pinning barrier for one wall type and decreasing it for the other. We believe that this effect explains the linear dependence of $\Delta B_{\text{dep}}$ on observed in Fig. 5 (b) for small in-plane fields. We also confirmed that the slope of this linear dependence scales with $\Delta H_{\text{DMI}}$ in the simulations (Fig. 6).

# Figures

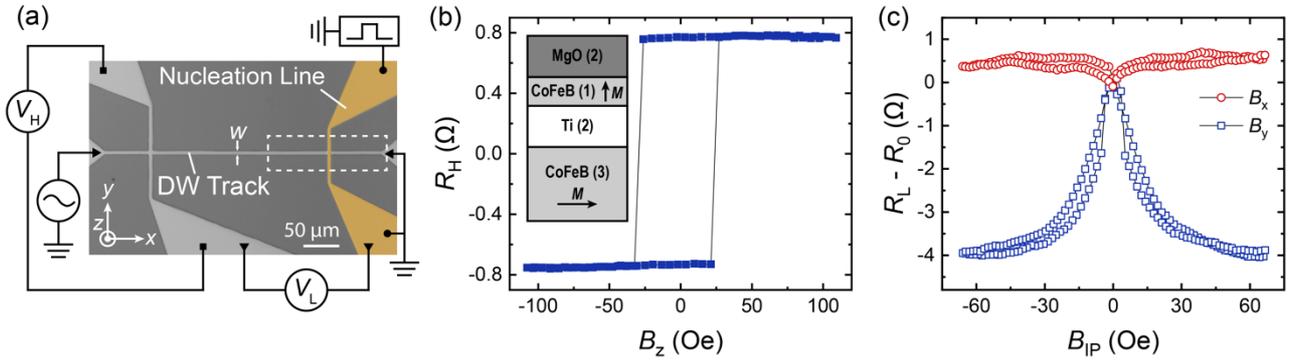

**Figure 1** - (a) Device micrograph, electrical connections and definition of coordinate system. The width of the DW track is $w = 5\ \mu$m. The Hall bar is used to characterize the magnetization of the bottom and top CoFeB layers. Differential MOKE images are taken in the dashed area. (b) $R_H$ measured during an OOP field sweep. $R_H$ is proportional to the magnetization of the top CoFeB layer. The inset in (b) shows the full layer structure. (c) Longitudinal resistance $R_L$ (offset by the resistance at zero field $R_0$) measured during an IP field sweep along the $x$ and $y$ axes. Due to the large perpendicular magnetic anisotropy of the OOP layer, the signal is predominantly related to the bottom CoFeB layer and it is proportional to the $x$-component of the magnetization through the anisotropic magnetoresistance.

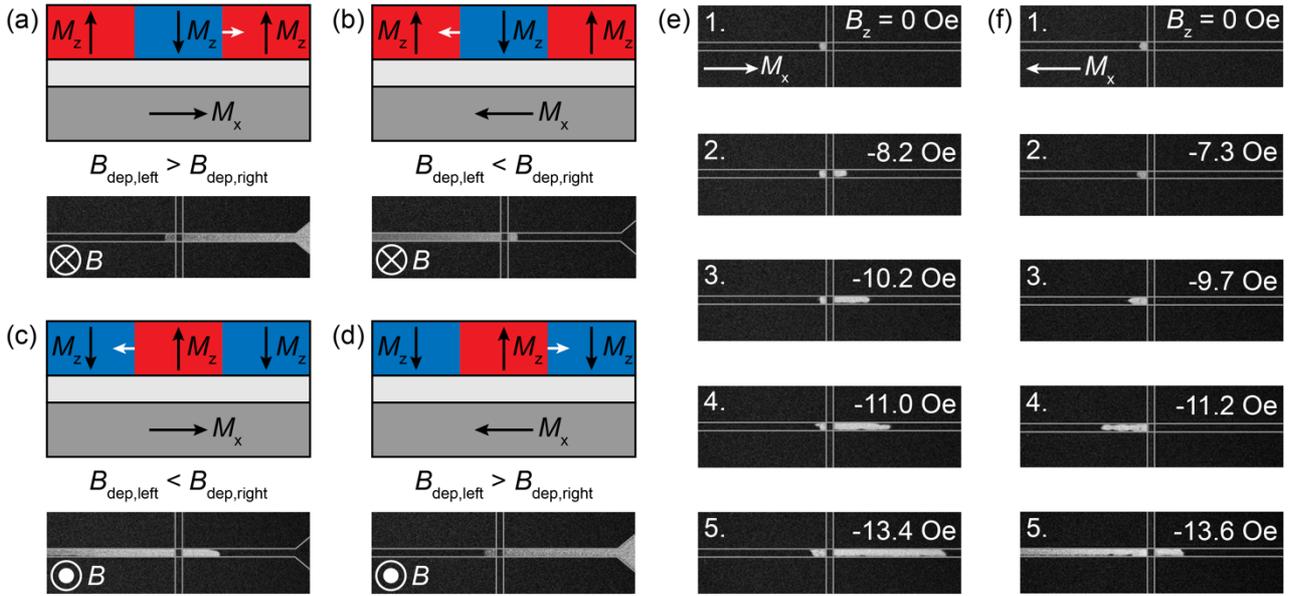

**Figure 2** - (a-d) The top panel shows a schematic side view of the initial magnetic configuration. Blue (down) and red (up) correspond to the magnetization of the top CoFeB layer. Black arrows correspond to the magnetization of the top and bottom CoFeB layers, the white arrow indicates the direction along which the DW will depin at a lower field. The bottom panel is a differential MOKE image taken during the ramping of $B_z$ after the first domain wall has depinned and moved along the track. Note that in (c) the second wall has also depinned. (e-f) Asymmetric expansion of a domain for the initial configurations shown in (a-b), respectively. The initial domain is nucleated using the Oersted field from a current pulse through the nucleation line (vertical bar). Before $B_z$ is ramped (i.e. in between panels 1. and 2.), a new reference image is taken to enable the accurate measurement of the depinning field from the differential MOKE images. In panel 2.-5., the initial domain shown in panel 1. is superimposed back onto the image, so that the entire domain is shown. Note that when the first DW has to pass underneath the Au wire, as in (e) the initial expansion of the domain occurs at a higher field. The field ramp is 0.1 Oe s$^{-1}$.

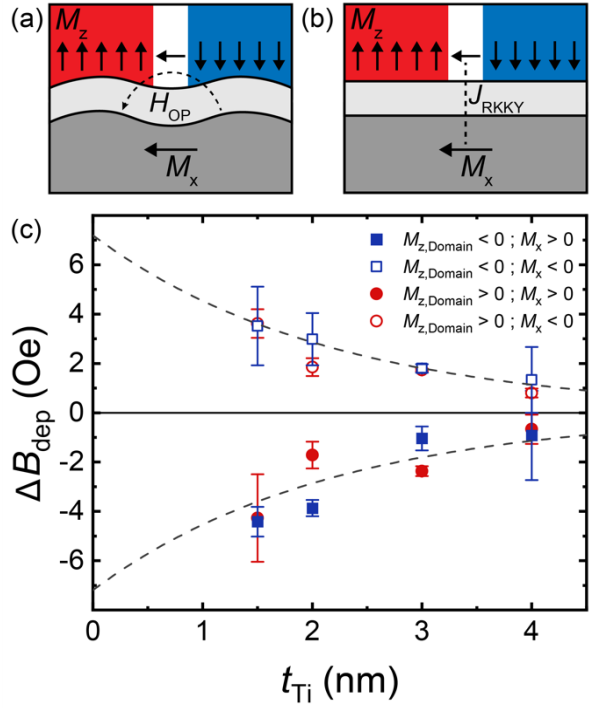

**Figure 3** – (a-b) Illustration of the magnetostatic coupling due to the 'orange-peel' effect and the exchange coupling due to the RKKY interaction. (c) Averaged difference between the depinning field ($B_{dep}$) of the left and right domain walls for all four configurations shown in Figure 2(a-d), plotted as a function of the Ti spacer thickness $t_{Ti}$. The error bars show the standard error of the averaged data. Dashed lines are a fit to the data (open blue squares) according to the exponential decay of the orange-peel effect. Note that for the sample with $t_{Ti} = 3$ nm, the measurement was only performed with one polarization of the nucleation pulse. The depinning field in this device was a few Oe for both DWs. Hence, saturating the IP magnetization after the nucleation pulse caused the DWs to move, due to a small misalignment. All data for $t_{Ti} = 3$ nm were obtained by nucleating a domain with the Oersted field parallel to the magnetization of the IP CoFeB.

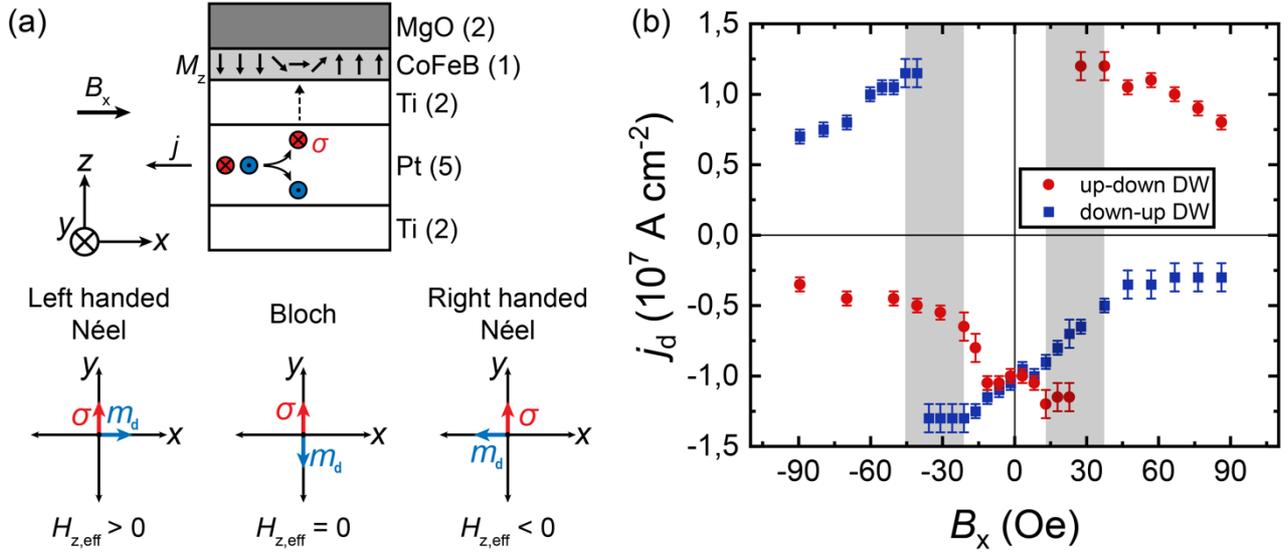

**Figure 4 -** a) Schematic of the layer structure used in the SOT-induced depinning experiments. σ is the polarization axis of the spins injected into the CoFeB. $m_d$ denotes the DW internal magnetization axis and $H_{z,eff}$ is the orientation of the SOT effective field acting on the DW. (b) Depinning current $j_d$ plotted as a function of $B_x$. The gray shaded areas represent the uncertainty interval for $\mu_0 H_{DMI}$. Note that a small bias field $B_z = 2$ Oe is applied to reduce the depinning current, independent of $B_x$.

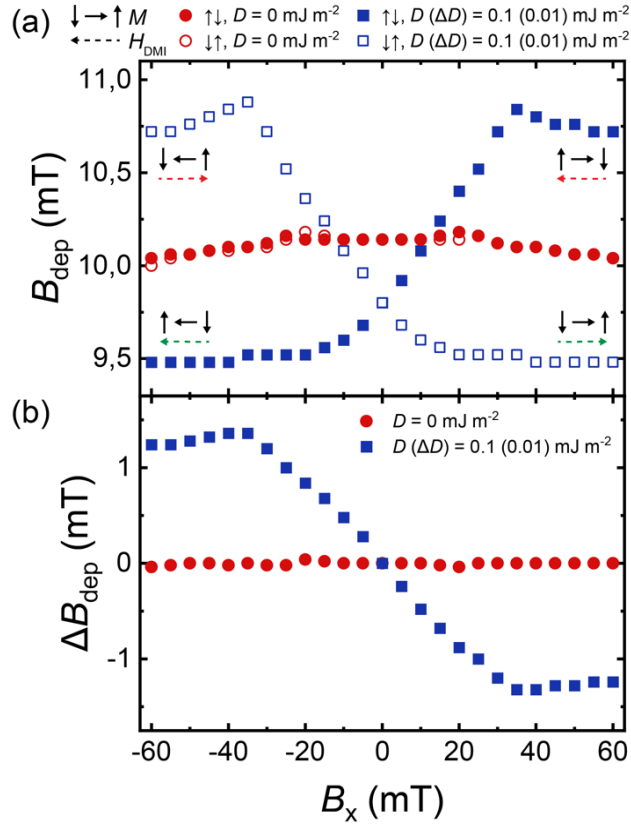

**Figure 5** – (a) Micromagnetic simulations of the DW depinning field ($B_{dep}$) as a function of the in-plane field ($B_x$) for four different scenarios: up-down (solid symbols) and down-up (open symbols) DWs in the absence (red dots) and presence (blue squares) of the DMI. The solid and dashed arrows indicate the the DW magnetic configuration and the DMI effective field $H_{DMI}$, respectively. (b) Calculated $\Delta B_{dep}(= B_{dep,left} - B_{dep,right})$ for a down domain as a function of $B_x$ in the absence (red dots) and presence (blue squares) of DMI. See text for further details.

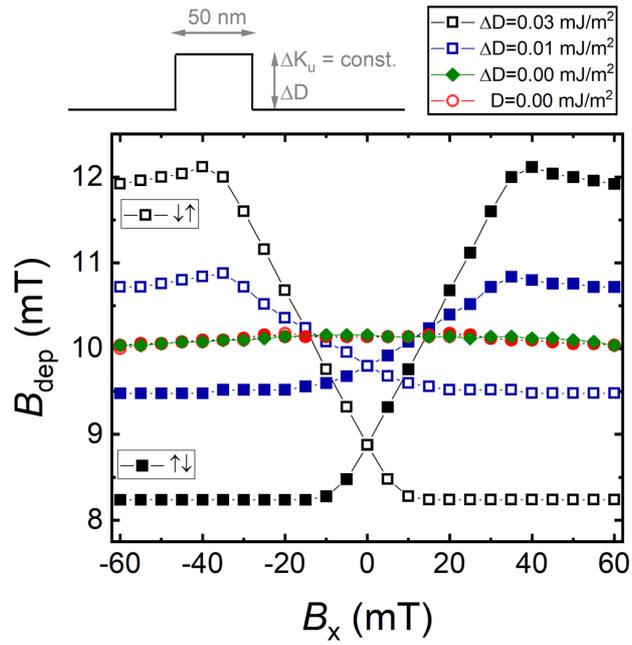

**Figure 6** - Simulation of the dependence of the depinning field on an in-plane magnetic field (representing an inter-layer coupling) for four different scenarios: no DMI (red), a uniform DMI of $|D| = 0.1$ mJ m-2 (green), an increased DMI in the pinning site given by $\Delta D$ (blue and black). Open or closed symbols correspond to up-down and down-up wall, respectively.